\input harvmac
\newcount\figno
\figno=0
\def\fig#1#2#3{
\par\begingroup\parindent=0pt\leftskip=1cm\rightskip=1cm\parindent=0pt
\baselineskip=11pt
\global\advance\figno by 1
\midinsert
\epsfxsize=#3
\centerline{\epsfbox{#2}}
\vskip 12pt
{\bf Fig. \the\figno:} #1\par
\endinsert\endgroup\par
}
\def\figlabel#1{\xdef#1{\the\figno}}
\def\encadremath#1{\vbox{\hrule\hbox{\vrule\kern8pt\vbox{\kern8pt
\hbox{$\displaystyle #1$}\kern8pt}
\kern8pt\vrule}\hrule}}

\overfullrule=0pt

\Title{MIT-CTP-2508, TIFR-TH/96-02}
{\vbox{\centerline{Excitations of D-strings, Entropy and Duality}}}
\smallskip
\centerline{Sumit R. Das\foot{E-mail: das@theory.tifr.res.in}}
\smallskip
\centerline{\it Tata Institute of Fundamental Research}
\centerline{\it Homi Bhabha Road, Bombay 400 005, INDIA}
\smallskip
\centerline{and}
\smallskip
\centerline{Samir D. Mathur\foot{E-mail: me@ctpdown.mit.edu}}
\smallskip
\centerline{\it Center for Theoretical Physics}
\centerline{\it Massachussetts Institute of Technology}
\centerline{\it Cambridge, MA 02139, USA}
\bigskip

\medskip

\noindent

We examine the BPS and low energy non-BPS excitations of the D-string,
 in terms
of open strings that travel on the D-string. We use this to study the
energy thresholds for exciting a long D-string, for arbitrary winding
number. We also compute the leading correction to the entropy from
non-BPS states for a long D-string, and observe the relation of all
these quantities with the corresponding quantities for the elementary
string.

\Date{January, 1996}
\def\TD{{T^{(D)}}}
\def\TS{{T^{(S)}}}

\def\GE{G^{(E)}}

\def\tpsi{{\tilde{\psi}}}
\def\vp{{\vec p}}
\def\vk{{\vec k}}

\newsec{Introduction}

Recently the idea that massive elementary particles behave as black
holes at strong coupling
\ref\salam{S. Hawking, {\it Month.
Not. Roy. Astro. Soc.}~{\bf 152} (1971) 75; A. Salam, in {\it
Quantum Gravity : an Oxford Symposium} (eds. C. Isham, R.Penrose
and D. Sciama, Oxford University Press, 1975); G. 't Hooft,
{\it Nucl. Phys.} ~{\bf B335} (1990) 138; J. Preskill, {\it 
Physica Scripta} {\bf T36} (1991) 258; C. Holzhey and
F. Wilczek, {\it Nucl. Phys.} ~{\bf B380} (1992) 447}
has proved to be very fruitful in string theory
\ref\susskind{L. Susskind, hep-th/9309145; J. Russo and L. Susskind,
{\it Nucl. Phys.}~{\bf B437} (1995) 611}
\ref\duff{M. Duff, M. Khuri, R. Minasian and J. Rhamfeld,
{\it Nucl. Phys.}~{\bf B418} (1994) 195; M. Duff and
J. Rahmfeld, {\it Phys. Lett.} {\bf 345B} (1995) 441}.
In particular, it has been argued that this could lead to an
understanding of black hole entropy \susskind.
In fact in
\ref\sen{A. Sen, {\it Nucl. Phys.}~{\bf B440} (1995) 421;
{\it Mod. Phys. Lett.}~{\bf A10} (1995) 2081; for an earlier
suggestion see C. Vafa as quoted in \susskind}
BPS black holes in
heterotic string theory compactified on $T^6$ were shown to have the
same entropy (defined as the area of the stretched horizon) as that
expected from the degeneracy of BPS states in the elementary string
theory.  The importance of BPS states lies in the fact that their
masses are not changed by quantum corrections and that a certain class
of these are absolutely stable. In fact the correspondence has been
understood \ref\wilc{F. Larsen and F. Wilczek, hepth/9511064.}
\ref\stromvafa{A. Strominger and C. Vafa, hepth/9601029}
in certain other BPS black holes in this theory
having nonzero horizon area \ref\cvetic{M. Cvetic and D. Youm,
hepth/9512127 and references therein.}.

Evidence in favor of the identification of elementary BPS states of
heterotic strings with extremal black holes has been also obtained in
scattering processes of these states in \ref\jerome{
J. Gauntlett, J. Harvey, M. Robinson and 
D. Waldram, {\it Nucl. Phys.}~
{\bf B 411} (1994) 461; R. Khuri and R.
Myers, hep-th/9508045} and
\ref\callan{C.G. Callan, J.M. Maldacena and A.W. Peet, hep-th/9510134}
(in \callan, as well as in
\ref\dabholkar{A. Dabholkar, J. Gauntlett,
J. Harvey and D. Waldram, hepth/9511053.} such black holes are
obtained by compactifying macroscopic strings \ref\dharvey{A.
Dabholkar and J. Harvey, {\it Phys. Rev. Lett.}~{\bf 63} (1989) 2073;
A. Dabholkar, G. Gibbons, J. Harvey and F. Ruiz-Ruiz, {\it Nucl.
Phys.}~{\bf B340} (1990) 33.}), and in scattering of massless scalar
states from such BPS states in
\ref\wadia{G. Mandal and S.R. Wadia, hepth/9511218}
(where the inelastic thresholds  are also examined).

If the above reasoning is correct other BPS states in string theory
like those
which carry RR charges should behave in a similar manner.  In fact Type IIB
string theory is conjectured to be self dual in ten dimensions. In
this theory there are macroscopic string solutions
\dharvey\ which are to be identified with the BPS states of the
elementary string carrying NSNS charge. An $SL(2,Z)$ symmetry which mixes
the NSNS and RR rank three gauge fields can be used to generate
other solutions which carry both NSNS and RR charges
\ref\schwarz{J. Schwarz, hepth/9508143} - in particular strings which
carry only RR charges. In a recent work, Polchinski 
\ref\polchinski{J. Polchinski, hepth/9510017}
has shown how to describe objects in string theory which carry RR
charges - through open strings with Dirichelt boundary conditions in
some of the directions, giving `D-branes' \ref\dbranes{J. Dai, 
R. Leigh and
J. Polchinski, {\it Mod. Phys. Lett.}~{\bf A4} (1989) 2073; 
P. Horava, {\it Phys. Lett.}~{\bf B231}~(1989) 251;
R. Leigh,  {\it Mod. Phys. Lett.}~{\bf A4} (1989) 2767; 
M. Green, {\it Phys. Lett.}~{\bf B266} (1991) 325 and
{\it Phys. Lett.}~{\bf B329} (1994) 435}
\ref\polchan{J. Polchinski, {\it Phys. Rev.}~{\bf D50} (1994) R 6041}. 
These are exact descriptions of RR
$p$-branes ( $p$ is odd for Type IIB and even for Type IIA).  The
D-brane descrption has been used to provide strong evidence for the
existence of these $SL(2,Z)$ multiplets \ref\witten{E. Witten,
hep-th/9510135},\ref\senbrane{A. Sen, hep-th/9510229 and 9511026; C.
Vafa, hep-th/9511088}.

In this paper we consider D-strings in the Type-IIB theory.  We
examine the set of open strings that can live on this D-string, and
thereby identify and count the BPS excitations of the D-string. We
also examine low energy non-BPS excitations in this manner and compute
their entropy - such excitations should be long lived if the D-string
is sufficiently long and the level of excitation small.  We also
consider the scattering of massless probes from a long D-string using
the results of \ref\klebanov{I. Klebanov and L. Thorlacius,
hepth/9510200} and show that at low energies the expected Coulomb form
is obtained.  Furthermore the threshold for inelastic scattering from
elementary string states \wadia\ is shown to match exactly with that
obtained with our identification of the non-BPS excited states.  In
all of the above it is interesting to note the distinction between the
D-string wound $n_w$ times and the case of $n_w$ singly wound
D-strings placed next to each other.

Backgrounds of such excitations have been analysed in
\ref\KLEBCALL{C.G. Callan and I. Klebanov, hepth/9511173}. Such
backgrounds, however, describe coherent states of excitations of the
D-string, while in the discussion below
we wish to count all excitations.

\newsec{ The macroscopic NSNS and RR Strings}

We will use the Einstein metric $\GE$ throughout this paper.

The low energy effective action of the type IIB theory is
\eqn\onep{S~=~{1 \over 2 \kappa^2} \int d^Dx~\sqrt{-\GE}~
[R~-~{1\over 2}(\nabla \phi)^2 ~-~{1\over 12}e^{-\phi}(H^{(1)})^2~-~
{1\over
12}e^{\phi}(H^{(2)})^2 ]} $H^{(1,2)}$ 
are the rank three NSNS and RR
field strengths respectively related to the corresponding gauge fields
$B^{(i)}$ by $H^{(i)} = dB^{(i)}$.  We have omitted the other fields
which are irrelevant for our purposes.  This action has a $SL(2,Z)$
symmetry which mixes the two rank two gauge potentials \schwarz. A
particular case of this duality is the transformation
\eqn\oneq{\phi \rightarrow -\phi~~~B^{(i)} \rightarrow 
\epsilon^{ij} B^{(j)}}

Suppose we add to the action \onep\ the source term from the
worldsheet of the elementary string
\eqn\four{S_{source} = {T \over 2}\int d^2\xi
 [(e^{\phi/2}\GE_{AB}(X)\delta^{ab}
+B^{(1)}_{AB}(X)\epsilon^{ab})
\partial_a X^A \partial_b X^B]}
 We take 10-dimensional spacetime with $X^9$ compactified on a
circle of circumference $L$. Then we have the following solution for
the low energy fields if the elementary string is wound $n_w$ times
around $X^9$ \callan, \dabholkar\
\eqn\five{\eqalign{ds^2 & = A^{-3/4}[-dt^2
 + dz^2] +A^{1/4} d{\vec x}. d{\vec x} \cr
A & = 1 + \sigma \Lambda,~~~~~~e^{-2\phi}~=~
e^{-2\phi_0}A,~~~~~~~~B^{(1)}_{90}= e^{{\phi_0 \over 2}}A^{-1}\cr
\Lambda &= {\kappa^2 \over 3 \omega_7} {1 \over r^6}
~~~~~\sigma = n_w Te^{\phi_0/2}}}
Here ${\vec x} = (x^1,\cdots x^8)$ and $r^2 = {\vec x}\cdot{\vec x}$ 
and $\omega_7$ is the volume of the unit seven sphere.
The ADM mass of the macrosopic string may
be obtained from the asymptotic form of $g_{00}$ and is given by
\eqn\nine{ M = \sigma L }

We can add transverse oscillations to the string which also carry a
corresponding amount of momentum along the string; such solutions are
also described in \callan \dabholkar . We will however be later
interested in oscillations of the string that are not necessarily
coherent classical deformations, and for these the classical fields
must be approximated by quantum and phase averages.

A string solution in lower dimensions may be obtained easily by
compactifying $10 - d$ of the transverse dimensions ${\vec x}$ on a
compact manifold of volume $V_c$ \callan, \dabholkar.  For distances
much larger than the size of the compact manifold one has a solution
which is identical to \five, with the sole change being in the form of
$\Lambda$ which now becomes $\Lambda_d$
\eqn\eight{\Lambda_d = {2\kappa^2 \over V_c (d-4) \omega_{d-3}}~
{1 \over \rho^{(d-4)}}} 
where $\rho$ is the radial distance in the
theory reduced to $d$ dimensions.

A string solution with RR charges may be now obtained using
the `duality transformation' \schwarz. For a string carrying purely RR
charge the solution is obtained from \five\ by reversing the sign of
the $\phi$ and $\phi_0$ and replacing $B^{(1)}$ by $B^{(2)}$
\foot{Note that the relation (17a) in \schwarz\ describing the
antisymmetric tensor field for general string solutions with both NSNS
and RR charges is not quite correct : the vector $B^{(i)}$ in that
equation has to be replaced by ${\cal M}_0^{-1}~B$ where ${\cal M}_0$
denotes the value of the matrix ${\cal M}$ at infinity.}.

\newsec{String states and macroscopic strings}

Our goal is to compare states of the elementary string carrying NSNS
charge with states of the dual string carrying RR charge. As in
section 2, we will consider the elementary string to be wound $n_w$
times around the compact direction $X^9$.  The mass of such states (in
Einstein metric) is given to lowest order in the coupling by
\eqn\oneone{\eqalign{m^2 &= (n_wL\TS+{2\pi n_p\over L})^2
+ 8\pi \TS(N_R-\delta_R)\cr
& = (n_wL\TS-{2\pi n_p\over L})^2+8\pi
\TS(N_L-\delta_L)}} 
where $\TS = T e^{{\phi_0 \over 2}}$ and $n_w, n_p$
are integers giving the winding and momentum in the $X^9$ direction,
and $\delta_{L,R} =0,1/2$ for the Ramond and Neveu-Schwarz sectors
respectively.  In the following we will consider very long strings, so
that $L{\sqrt{\TS}} >> 1$.

The D-string, on the other hand, appears as a solitonic string with
tension $\TD=Te^{-\phi_0/2}$.  Open strings have Dirichlet boundary
conditions on the D-string:
\eqn\fifteen{\eqalign{ (\alpha_n^\mu~+~\tilde\alpha_{-n}^\mu)|B>~=
&~0,~~(\psi^\mu_n -i\tpsi^\mu_{-n}) |B>~=~0, ~~ \mu=0,9\cr
(\alpha_n^i~-~\tilde\alpha_{-n}^i)|B>~=&~0,~~(\psi^i_n +i
\tpsi^i_{-n}) |B>~=~0, ~~ i=1\dots 8\cr }} 

We will examine the excitations of the D-string by studying
configurations of open string states that can live on the D-string.
Due to Dirichlet boundary conditions in the eight transverse
dimensions the open string states can have nonzero momenta only along
the longitudinal directions $X^0$ and $X^9$. The lowest mass states of
the open string are the massless excitations
\eqn\onetwo{\eqalign{\Psi_b~=&~\psi^i_{-1/2}|p>,~~p_0
=|p_9|,~~p_i=0, ~~(i=1\dots 8)\cr
\Psi_f~=&~|p>_{\alpha},~~p_0=|p_9|,~~p_i=0,  ~~
(i=1\dots 8,~~\alpha=1\dots 8)\cr}} 
Here the two classes of states
come from the NS and R sector of the open string respectively.
$\alpha$ is the spacetime spinor index; we are in the light cone gauge
for the open string where both vector and spinor indices run over $8$
possibilities.

Apart from the states in \onetwo\ there are open string states which
involve higher oscillator content. We will not consider them since in
general these states will decay at strong coupling.\foot{We thank M.
Douglas for pointing this out to us.}

For the discussion below we use the term `ground state' of the
elementary string for the state having some nonzero winding number
$n_w$ but $n_p = 0$, $N_R=\delta_R$, $N_L=\delta_L$. Correspondingly,
we take the `ground state' of the D-string to be one with no
transverse excitations and no $B^{(1)}$ charge.

\subsec{Ground States}

Consider the elementary string with $n_w=1$, in the `ground state' as
described above. This state has a degeneracy equal to $16 \times
16=256$, where the $16$ states in each of the left and right sectors
come from the $8$ bosonic ground states in the NS sector and the $8$
fermionic ground states in the R sector. Thus there are $128$ bosonic
and $128$ fermionic `ground states' of the elementary string.

For the D-string, we consider the description of \witten\ where the
low energy field theory coming from the open strings is the
dimensionsal reduction to the 2-dimensional D-string world sheet of
the 10 dimensionsal supersymmetric U(1) gauge theory. This theory has
8 scalars corresponding to the transverse deformations of the
D-string, and 8 majorana spinors on the 2-dimensional space.  The
latter give 8 zero modes, which generate $2^8=256$ degenerate ground
states of the supersymmetric field theory.

\subsec{BPS states with $n_w = 1$}

There are an infinite set of BPS saturated states in the fundamental
Type II string. These are states with either (i) $N_R = \delta_R$ with
arbitrary $N_L$ or (ii) $N_L = \delta_L$ with arbitrary $N_R$ and with
arbitrary values of $n_p$ and $n_w$ in either case \susskind\ 
\duff\ \sen . For such
states the mass formula \oneone\ is exact.  We will choose $N_R =
\delta_R$ and also restrict for the moment to $n_w =1$. One therefore
has, from
\oneone, $N_L = \delta_L + n_p$ ($n_p>0$). The mass is
\eqn\twoone{m = {\bar m}_{NSNS} + {2 \pi n_p \over L},~~~~
~~~~  {{\bar m}_{NSNS} = \TS L}}

 The degeneracy of such
states is given by the number of ways one can decompose $n_p$ into
levels of the $8$ bosonic and $8$ fermionic oscillators.  As is well
known, for large $n_p$ the degeneracy behaves as an exponential of
${\sqrt{n_p}}$. This leads to an entropy \susskind\ \sen\
\eqn\oneeight{S~\sim~2\pi\sqrt{cN_L/6}~=~2\pi\sqrt{2n_p}}
where we have used the fact that the central charge
$c$ comes from 8 transverse bosons
and 8 transverse fermions and thus totals 12.  This should be
identified with the ``black string'' entropy.  One expects that the
black string entropy is proportional to the length of the string. This
is indeed so if we rewrite the answer in terms of the parameters
which appear in the corresponding classical solution.

We identify the corresponding states of the D-string as those that
have open strings in the states of the form \onetwo , with
\eqn\qone{p_9~=~{2\pi m\over L}, ~~~m>0, ~~~~\sum m = n_p}
The mass is
\eqn\twothree{m = {\bar m}_{RR} + 
{2 \pi n_p \over L}, ~~~~~~~~{{\bar m}_{RR} = \TD L}} 
In the low
energy limit such a state has the metric corresponding to the string
tension ${\bar m_{RR} \over L}$ and momentum $P=2\pi n_p/L$.

The degeneracy of such D-string states is given by the number of ways
one can decompose the integer $n_p$ into the positive integers $m$'s
of the individual open string states.  Note that there are $8$ bosonic
and $8$ fermionic single string states of the type \onetwo\ for a
given momentum. It is immediate that we get the same combinatorics as
for the elementary string, where the {\it oscillator levels} on the
left sector were positive integers that had to total upto $n_p$ to
generate all the BPS states. Thus we get the same entropy \oneeight\
at the same total momentum $P=2\pi n_p/L$ for the D-string as for the
elementary string, in accordance with duality.

To describe the BPS states corresponding to with $N_L=0$, $N_R\ne 0$
 we take the
momentum $p_9$ of each open string to be negative instead of positive.

We expect these excited states of the D-string to be stable because if
for instance a pair of open strings were to decay to a closed string,
by conservation of $p_0$ and $p_9$ the closed string would have no
transverse momentum $p_i$.  ($p_0^2\ge p_9^2+p_i^2$; but $p_0=p_9$
from the open strings.)  If there are noncompact directions among the
$X^i$ then the overlap of the open strings (which are confined to the
neighbourhood of the D-string) and the zero $p_i$ state of the closed
string vanishes, so there is no phase space for decay.

\subsec{Non-BPS states with $n_w = 1$}

We now consider the non-BPS states specified by some momentum $P=2\pi
n_p/L$ ($n_p>0$) and $n_w = 1$. For elementary string states it 
follows from
\oneone\ that these are described by
\eqn\twofive{N_R = \delta_R +  n,~~~~~~N_L = \delta_L + n_p + n,
~~~~(n = {\rm integer})}
The mass (at lowest order in coupling) for such a state $m_n$ is
given by 
\eqn\threeone{m_n = \sqrt{[{\bar m}_{NSNS}+ (2\pi |n_p|/L)]^2 
+ 8\pi \TS n}} 
where $\bar m_{NSNS}$ is given in \twoone. For large $L{\sqrt \TS}$ 
one has
\eqn\twosix{m_n = {\bar m}_{NSNS}+ {2\pi |n_p|\over L} + 
{4 \pi n \over L} + O({1 \over \TS L^3})}
In particular the lowest mass state of this kind has $n_p = 0$ and $n
= 1$. The degeneracy of such states comes from the many possibilities
for the oscillator excitations, in both the left and the right sectors
independently. We thus have, for large $n$ the total degeneracy
behaving as
\eqn\twoseven{D(n) \sim e^{2\pi{\sqrt{2(n_p + n)}}}~e^{2\pi
{\sqrt{2n}}}} 
This gives the entropy for the non-extremal string. For
example for $n_p = 0$ one has an entropy
\eqn\twoeight{S \sim 4\pi{\sqrt{2n}} = {\sqrt{[4\pi(m_n^2 - {\bar m}
_{NSNS}^2]/\TS]}}}

To construct the corresponding non-BPS states for the D-string we let
there be open strings on the D-string with both signs of $p_9=2\pi
m/L$, with $\sum m=n_p$ as before but with $\sum |m|=|n_p|+2n$. The
mass of such a state is given by
\eqn\twonine{m_n ~\approx~ {\bar m}_{RR} + {2\pi |n_p|\over L}
+ {4 \pi n \over L}}

Let us analyse the approximation in the above equation.
 A pair of oppositely moving open
strings scatter through a disc diagram to another pair of open
strings. This diagram can generate shifts in the energy of a state of
the excited D-string.  Let all the four open strings have momenta
 $p_9=\pm 2\pi n/L,~p_0\equiv\omega=
|p_9|$. Then the amplitude for scattering per unit time is
$ \sim e^{\phi_0} \TS (L\omega)^{-2}L (\omega^2/\TS)^2 \sim
(\TD n^2 L^3)^{-1}$.  We expect
 energy shifts for the states of this  order,
which is the correct order to be dual to the term dropped in \twosix .
(In this amplitude calculation one should note that the leading 
terms in $\omega$
 in the 4 point disc amplitude cancel after summing over all
 the cyclically inequivalent orderings of vertex opertors.)

We now wish to count non-BPS excitations of the D-string (at large
coupling $e^{\phi_0}$) and see if they agree with the non-BPS
excitations of the elementary string (at small coupling). As
mentioned above, the masses of non-BPS states on the D-string will get
corrections due to open string scattering to open strings.  Further,
there is a process where two open strings travelling in opposite
directions interact and decay into a closed string.  We take our
limits in the following way. First we take the coupling sufficiently
strong so that only the states
\onetwo\ are the stable open string states; it is reasonable to
expect that the
higher oscillator mode open strings decay rapidly and can be
ignored. We fix a certain upper
limit for the integers
$n,~n_p$ in \twonine ; we will count states
upto these excitation numbers. Then we take the limit $L$ large.
In this limit it becomes hard for the open strings to
`find' each other and interact, 
 so that both the energy corrections to the non-BPS states
and the rate of decay of these states (to closed strings)
go to zero. Then we can reliably count these non-BPS states.

 But now we
immediately see that this count is again identical to that in the
elementary string case since there is a one-to-one correspondence
between open strings with positive and negative $m$ on the one hand
and the oscillator mode numbers in the right and left sectors of the
Type II elementary string on the other. We therefore get the same
expression \twoseven\ for the degeneracy and thus
for  the entropy of `slightly nonextremal' states of the
D-string.

\subsec{ States with $n_w > 1$}

Now consider the case $n_w>1$. This generalization is trivial for the
elementary string states since everything follows from the mass
formula
\oneone. For BPS states the expressions \twoone\ are modified, giving
 $m = n_w L\TS + {2\pi n_p \over L}$,  
$N_L = \delta_L + n_w n_p$,  leading to an entropy 
$S \sim 2\pi{\sqrt{2n_w n_p}}$. 

In the case of the D-string we need an ansatz for the behavior of the
open string momenta for the case $n_w>1$. In particular this ansatz
must distinguish between the case (a) where we have $n_w$ different
D-strings, each with winding number unity, next to each other, and the
case (b) where we have a single D-string that has winding number
$n_w$. To see that a distinction is required note that for the
analogous cases in the elementary string, the excitation energies and
degeneracies are different for these two cases.

For the case (a) where we have $n_w$ different singly wound D-strings
it is sufficient to let the open strings have momenta $2\pi m/L$ as in
the discussion of sections 3.2, 3.3, but with a Chan-Paton factor at
each end of the open string which takes $n_w$ values corresponding to
which of the D-strings the end is on; this gives an $n_w^2$ fold
degeneracy to the excitations for a single
D-string \polchan\ \witten .

For the case (b) of the single D-string with winding number $n_w$ the
following ansatz appears to be adequate. We postulate that each single
open string state can
 carry momentum $p = {2\pi m \over n_w L}$ ($m$ an integer) but the
total momentum carried by all the open strings has to equal ${2\pi n_p
\over L}$, $n_p$ an integer.
 Thus the translation of the entire system in the compact
direction through $L$ returns the wavefunction of the system to itself
with no phase.

It is easy to see that with this ansatz we reproduce the number of BPS
and non-BPS states that we find from the elementary string, by
carrying out the steps similar to those in the above subsections.
In particular for BPS states we have to partition
 the integer $n_w n_p$ among positive integers, just as for the
 elementary string.  To
see why this ansatz for the momenta is reasonable, note that the open
strings describe oscillations of the D-string which now has length
$n_w L$; thus they must have correspondingly long wavelengths (i.e. low
energies).

In particular let us check the lowest energy excitation having zero
total momentum. The elementary string has for such a state the mass
${4\pi\over n_w L}$ (coming from the obvious modification of \twosix\
). This agrees with the energy of for the lowest energy excitation of
the D-string: a pair of open string states with momenta $\pm
{2\pi\over n_w L}$ and hence carrying total energy ${4\pi\over n_w
L}$.

In the description of \witten\ we have a supersymmetric $U(n_w)$ Yang
Mills theory on the D-brane. On the basis of the above discussion we
would expect that for the supersymmetric vacuum corresponding to a
single long D-string, the lowest excitations should have energies
$E={4\pi\over n_w L}$, momentum $P=0$. This energy is lower (for
$n_w>2$) than the energy of Goldstone modes in a periodic box of
length $L$.

\newsec{Scattering from string states and excitation thresholds}

In a recent paper \wadia\ the threshold for exciting
a BPS state of the elementary string to the lowest non-BPS state with
the same charges was computed.  In view of the conjectured duality the
 scattering
of probes from the D-string should agree with that from elementary
string states.

First we note that the elastic scattering cross section for neutral
scalars at low energies must be the coulomb scattering obtained from
the classical gravitational field of the string.
\foot{A discussion of low energy elastic  scattering  
has also been presented recently in \ref\KLEBTWO{S.S.
Gubser, A. Hashimoto, I.R. Klebanov and J.M. Maldacena, hepth/
9601057}, where scattering of RR scalars and polarisation dependences
have also been studied. } 

 We work in a
situation where five of the transverse directions $x^i~, i = 4,\cdots
8$ are compactified on an internal manifold of volume $V_c$.  We
consider scattering of massless gravitons of the closed string whose
polarizations $\epsilon_{ij}$ are nonzero only for $i = 4, \dots 8$
and whose momenta are nonzero only along the four noncompact
directions, i.e. $p^i \neq 0$ for $i = 0,\cdots 3$.  The target
(D-string or elementary string state) is taken to be in the lowest
state for the given winding.

The scattering cross section from a NSNS state of mass $M$ may be in
fact read off from the scattering in the heterotic string theory
computed in \wadia , by specialising to the case where the charges
come entirely from winding and momentum in the string direction. The
differential cross section in the rest frame of the target has the
expected Coulomb behavior at low energies :
\eqn\sixteen{ {d\sigma \over d\Omega} \sim (G_N M)^2~{\rm cosec}^4
{\theta \over 2}} where $G_N$ is the Newton's constant in 4 dimensions
and $M$ is the mass.  In \sixteen\ $\theta$ is the angle between the
initial (spatial) momentum $\vp$ and the final momentum $\vk$.  As
shown in \wadia\ this cross section agrees with that obtained from
classical scattering of a massless wave from a black hole of mass $M$.
The threshold for inelastic scattering corresponds to excitation of a
non-BPS state of the type \threeone\ with $n = 1$, $n_p=0$.  Simple
kinematics leads to the threshold energy \wadia
\eqn\fourthree{p^{th}_0 \approx {4\pi \over n_w L}}


The scattering from a fixed D-string can be computed by evaluating the
two point function of our ``graviton'' operators on a disc with
boundary conditions appropriate for D-strings. This calculation has
been already performed in \klebanov\ and we will simply use their
result. The amplitude for this process is
\eqn\twenty{{\cal T} \sim n_w \kappa^2 \epsilon^{(1)}_{ij}
\epsilon^{(2)}_{ij}
s~{\Gamma(1 - {s \over 2\pi T})
\Gamma(-{t \over 8\pi T}) \over \Gamma(1 - {s \over 2\pi T}
-{t \over 8\pi T})}}
where for our process we have
\eqn\twentyone{ s = k_0^2 - k_9^2 = p_0^2 - p_9^2,~~~~
 t = -(p+k)^2 = 2(k_0^2+
\vp \cdot \vk)}
and we have used the momentum conservation for the longitudinal
directions $p_0 = - k_0$, $p_9 = - k_9$ and the mass shell conditions
$k_0 = |\vk|,~~p_0 = |\vp|$. The factor of $n_w$ as compared to the
formulae in \klebanov\ follows from the fact that a D-string with
winding number $n_w$ has $n_w$ units of RR charge.  In the low energy
limit $k_0 \rightarrow 0$ the matrix element approaches the standard
Coulomb behaviour and one gets a cross section
\eqn\twentythree{{d\sigma \over d \Omega} ~
\sim ~ ({e^{-\phi_0}\TS n_w\kappa^2 \over V_c})^2 {\rm cosec}^4~{\theta
\over 2} =~(\TD n_w L \kappa_4^2)^2 {\rm cosec}^4~{\theta
\over 2}}
which agrees with the dependence on parameters with \sixteen\ since
$\TD n_w L=M$ and $G_N=8\pi \kappa_4^2$.
 ($\kappa_4$ is the  effective four dimensional coupling.)

The threshold for inelastic scattering may now come from any of the
following sources.

(a) \quad First, the D-string can be excited to the lowest non-BPS
state described in the previous section, i.e. a pair of open string
modes in their lowest allowed oscillator state and with equal and
opposite longitudinal momenta $\pm {2\pi \over n_w L}$.  The threshold
energy, ${4\pi \over n_w L}$ matches with the result \fourthree\ for
the elementary string. This is the lowest threshold for large
$L$.

(b) \quad One can excite states of the open string which have higher
number of oscillators. The threshold here is independent of $L$ and
equal to ${\sqrt{8\pi\TS}} = \sqrt{8\pi} T^{1/2} e^{\phi_0 \over 4}$.
This is the relevant threshold for small $L$ but will be a mild
resonance at strong coupling since these do not describe stable
states.  This kind of excitation is considered in \KLEBTWO\ for 
0-branes representing black holes.

(c) \quad One can excite a pair of open string states which wind in
opposite senses around the compact directions $x^i~,~i=4,\cdots 8$.
The corresponding threshold energy is $2\TS a=2Tae^{\phi_0 \over 2}$,
 where
$a$ is the smallest circumference of compactification. When $L >> {1
\over \TS a}$ this threshold is higher than the first one discussed
above.

None of these thresholds appears to represent the classical threshold
that would folllow from extending the calculation of \wadia\ to the
long black string. In fact the details of that calculation depend only
on the metric in the directions perpendicular to the string, and the
result is simply that the threshold is at wavelengths
\eqn\fouroneq{\lambda\sim {\kappa_5^2 M\over L}= 
{\kappa^2 \over V_c}Te^{-\phi_0/2} n_w ~
\sim~{1\over V_c}e^{3\phi_0/2} T^{-3} n_w}
where we have used the mass per unit length of the D-string.  Since
this thresold does not depend on $L$ it cannot be of type (a) above.
The threshold (b) above differs in its $T, \phi_0$ dependence from
\fouroneq . The threshold of type (c) above depends on the smallest
compact dimension; while \fouroneq\ depends only on the total volume
of all the compact dimensions $V_c$.

\newsec{Discussion}

We have observed the correspondence between excitations of the
elementary string and configurations of open strings travelling on
the D-string. The latter were described in a 
manner reminiscent of the Green-Schwarz language for
the elementary string.
 In particular we  estimated the entropy from
 non-BPS states
for a long D-string. Even for a  large  (but fixed) value of the
coupling, we could take  the length
$L$ sufficiently large so that the density of
non-BPS excitations is very low and thus  the decay rate and mass
corrections  for the non-BPS states can be ignored.
These long lived excitations  can be counted in a manner
 similar to the count of
BPS states.   The above approach to estimating non-BPS
entropy should have wider validity. We found that as we go away from
the extremal limit the leading correction to the entropy goes as
$\sim\sqrt{m-m_{\rm ex}}$ where $m-m_{\rm ex}$ is the mass in excess
of the BPS mass.  In a recent work the entropy of five dimensional
extremal holes with both electric and magnetic charge was shown to
equal the area of the horizon as given in \ref\GIBB{ G. Gibbons and K.
Maeda, {\it Nucl. Phys.}~{\bf B298} (1988) 741}. If
we take a black string in six dimensional space-time then we can apply
the above estimates. The correction to the horizon area in
\GIBB\ for slightly nonextremal holes indeed goes like
$\sim\sqrt{m-m_{\rm ex}}$, in accordance with the above expectation.

It would also be interesting to compute the amplitude for a D-string
excited in the above non-BPS states to fall into a black hole made
from other D-strings, and to see if the non-BPS excitations escape as
`Hawking radiation' in this process.

\newsec{Acknowledgements}

We have benefitted from discussions with A. Belopolsky,
 M. Douglas, M. Li, W. Taylor,
S.P Trivedi, S. R. Wadia and B. Zwiebach. S.R.D. would like to thank
the Theory Group of KEK for hospitality during the completion of this
paper. S.D.M. is partially supported under DOE cooperative agreement
number DE-FC02-94ER40818.

\listrefs
\bye